\documentclass[conference,compsoc]{IEEEtran} 
\IEEEoverridecommandlockouts
\usepackage{cite}
\usepackage{amsmath,amssymb,amsfonts}
\usepackage{graphicx}
\usepackage{textcomp}
\usepackage{xcolor}
\def\BibTeX{{\rm B\kern-.05em{\sc i\kern-.025em b}\kern-.08em
    T\kern-.1667em\lower.7ex\hbox{E}\kern-.125emX}}
\usepackage{url}

\usepackage{tabularx} 
\usepackage{natbib}
\usepackage{amsmath,amsfonts}
\usepackage{graphicx}
\usepackage{textcomp}
\usepackage{multirow}
\usepackage{multicol}
\usepackage{makecell}
\usepackage[dvipsnames]{xcolor}
\usepackage{amsthm}
\usepackage{xspace}
\usepackage{booktabs}
\usepackage[ruled,noend]{algorithm2e}
\usepackage{algpseudocode}
\usepackage{booktabs}
\usepackage{multirow}
\usepackage{siunitx}  
\usepackage{xcolor}   
\usepackage[table]{xcolor}  

\usepackage{booktabs}
\usepackage{multirow}
\usepackage{colortbl}
\usepackage{xcolor}
\usepackage{graphicx}
\usepackage{pgf}          

\def\BibTeX{{\rm B\kern-.05em{\sc i\kern-.025em b}\kern-.08em
    T\kern-.1667em\lower.7ex\hbox{E}\kern-.125emX}}

\newcommand{\sys}{\textsf{PRTESLA-C}\xspace}

\newcommand{\cellLow}[3]{%
  \pgfmathparse{abs(#3-#2) < 0.00001 ? 1 : max(min((#3-#1)/(#3-#2),1),0)}%
  \let\ratio\pgfmathresult
  \pgfmathparse{0.96 - 0.18*\ratio} \let\myred\pgfmathresult
  \pgfmathparse{0.78 + 0.20*\ratio} \let\mygreen\pgfmathresult
  \pgfmathparse{0.78}               \let\myblue\pgfmathresult
  \edef\temp{\noexpand\cellcolor[rgb]{\myred,\mygreen,\myblue}}%
  \temp #1%
}

\newcommand{\cellHigh}[3]{%
  \pgfmathparse{abs(#3-#2) < 0.00001 ? 1 : max(min((#1-#2)/(#3-#2),1),0)}%
  \let\ratio\pgfmathresult
  \pgfmathparse{0.96 - 0.18*\ratio} \let\myred\pgfmathresult
  \pgfmathparse{0.78 + 0.20*\ratio} \let\mygreen\pgfmathresult
  \pgfmathparse{0.78}               \let\myblue\pgfmathresult
  \edef\temp{\noexpand\cellcolor[rgb]{\myred,\mygreen,\myblue}}%
  \temp #1%
}

\begin{document}

\title{Tick-Tock on the Open Fronthaul: Securing Synchronization in O-RAN}

\makeatletter
\newcommand{\linebreakand}{%
  \end{@IEEEauthorhalign}
  \hfill\mbox{}\par
  \mbox{}\hfill\begin{@IEEEauthorhalign}
}
\makeatother

\author{
\IEEEauthorblockN{Yiwei Zhang}
\IEEEauthorblockA{Purdue University\\
West Lafayette, Indiana, USA\\
yiweizhang@purdue.edu}
\and
\IEEEauthorblockN{Enrico Pisanti}
\IEEEauthorblockA{Purdue University\\
West Lafayette, Indiana, USA\\
enrico.pisanti8.ep@gmail.com}
\and
\IEEEauthorblockN{Imtiaz Karim}
\IEEEauthorblockA{The University of Texas at Dallas\\
Richardson, Texas, USA\\
imtiaz.karim@utdallas.edu}
\linebreakand
\IEEEauthorblockN{Subangkar Karmaker Shanto}
\IEEEauthorblockA{Purdue University\\
West Lafayette, Indiana, USA\\
sshanto@purdue.edu}
\and
\IEEEauthorblockN{Elisa Bertino}
\IEEEauthorblockA{Purdue University\\
West Lafayette, Indiana, USA\\
bertino@purdue.edu}
}

\maketitle

\begin{abstract}
The Precision Time Protocol (PTP) provides the time and phase synchronization required by disaggregated Open Radio Access Networks (O-RAN). Yet, in current open fronthaul deployments, PTP traffic lacks mandatory authentication and integrity protection, leaving synchronization vulnerable to spoofing, replay, and delay manipulation attacks that can degrade radio access performance. Existing protections are poorly suited to this setting: they either add excessive latency, do not support multicast dissemination efficiently, or fail to contain key exposure under partially trusted RUs.

This paper analyzes the security risks of unprotected O-RAN PTP and develops a threat model for open fronthaul deployments. We then introduce \sys, a lightweight synchronization protection mechanism that combines per-round delayed key disclosure with ASCON-based message authentication. \sys uses an apply-then-verify-and-correct paradigm: timing samples are applied immediately to preserve real-time control, verified after key disclosure, and removed from persistent synchronization state if authentication fails. This design maintains sub-microsecond synchronization accuracy, provides strong protection against spoofing and replay, and bounds the impact of delay manipulation with minimal computational and latency overhead.

\end{abstract}


\section{Introduction}
The increasing demand for flexibility, scalability, and rapid innovation in 5G networks has accelerated the transition toward open and modular architectures. 
Open Radio Access Network (O-RAN) embraces this vision through a disaggregated design that separates hardware and software components, enabling programmability and multi-vendor interoperability~\cite{polese2023understanding,oran_3gpp_split,oran_planes,habibi2021towards}. 
However, this architectural openness also enlarges the attack surface. In particular, the Open Fronthaul (O-FH) interface between the O-RAN Distributed Unit (O-DU) and Radio Unit (O-RU) introduces new and insufficiently explored security risks~\cite{hung2024security,quad2023oransecurityreport,klement2024securing,nokia_oran_sec,atalay2023securing,thimmaraju2024security}.

Among the functions carried over the O-FH, time and frequency synchronization are especially critical~\cite{oran_planes,maamary2024synchronization,municio2023ran}. 
Accurate synchronization between O-DU and O-RU underpins key 5G features such as Time Division Duplex (TDD), beamforming, and coordinated multipoint transmission (CoMP). This functionality is typically realized through the IEEE 1588v2 Precision Time Protocol (PTP)~\cite{ptp1}. 
Yet, PTP was originally designed for controlled wired industrial environments, not for open, packet-switched, and potentially adversarial settings like O-RAN. When deployed over O-FH, PTP inherits security assumptions that no longer hold.

Despite its critical role, PTP lacks mandatory and robust mechanisms for message authenticity and integrity protection in current O-RAN deployments~\cite{polese2023understanding,dik2023open}. 
Existing specifications~\cite{oran_security_req,hirschler2011validation} often treat authentication as optional due to concerns about computational overhead and stringent real-time constraints. 
Consequently, synchronization traffic remains vulnerable to spoofing, replay, and delay manipulation attacks~\cite{dik2023open,oran_threat}, which can significantly degrade synchronization accuracy or even disrupt service availability.

Securing PTP in O-RAN is particularly challenging for three reasons. 
First, O-FH synchronization often relies on broadcast or multicast dissemination, especially when groups of O-RUs must maintain tight phase alignment, making pairwise secure channels impractical. 
Second, synchronization messages are exchanged at high frequency and directly feed real-time control loops, leaving little tolerance for added latency or jitter. 
Third, O-RUs may be partially compromised and leak locally stored credentials. Long-lived symmetric keys therefore become future-forgery capabilities: group keys amplify compromise across all O-RUs, while pairwise keys confine but do not eliminate the exposure.
Conventional solutions such as IPsec, TLS, and MACsec~\cite{dik2023open,mizrahi2011time} provide strong protection for unicast traffic, but their reliance on pairwise keying, heavyweight cryptographic processing, and limited multicast support render them ill-suited for high-frequency PTP flows under adversarial O-FH conditions. 
These challenges call for a lightweight, scalable, and timing-aware authentication mechanism tailored to O-RAN synchronization.

In this paper, we address these challenges by developing a security framework that aligns authentication strategies with O-RAN deployment constraints and realistic threat models. 
Inspired by the delayed authentication paradigm of TESLA~\cite{perrig2003tesla}, we revisit the design space of synchronization protection along three dimensions: key disclosure granularity, synchronization computation semantics, and verification timing. 
This analysis reveals a fundamental tension between immediate authentication (which preserves security but disrupts timing determinism) and delayed verification (which preserves real-time behavior but risks transient exposure).

Guided by this design space, we propose \sys (Per-Round TESLA with Correction), a scalable authentication scheme for securing PTP synchronization in O-RAN. 
\sys combines per-round delayed key disclosure with lightweight ASCON-based message authentication~\cite{dobraunig2021ascon}, and adopts an apply-then-verify-and-correct paradigm. 
Synchronization updates are applied immediately upon reception to maintain real-time responsiveness, verified once the corresponding key is disclosed, and corrected in a bounded manner if authentication fails. 
This approach preserves synchronization stability while preventing long-term impact from spoofed or manipulated messages.

\sys is built on three key principles. 
First, it eliminates long-lived shared group keys at O-RUs. 
By leveraging one-way key chains with delayed disclosure, \sys prevents a compromised O-RU from forging future synchronization traffic, mitigating compromise amplification risks inherent in symmetric-key designs. 
Second, it naturally supports multicast dissemination, enabling scalable protection for groups of O-RUs that must remain tightly synchronized. 
Third, it decouples authentication from real-time control while ensuring bounded correction, thereby maintaining sub-microsecond synchronization precision without destabilizing the control loop. 
Through lightweight cryptographic operations and efficient implementation techniques, \sys introduces minimal computational and bandwidth overhead.

In summary, our contributions are as follows:
\begin{itemize}
\item We present a comprehensive threat model for PTP synchronization in O-RAN, identifying attack vectors arising from unprotected fronthaul synchronization and analyzing the limitations of existing protection mechanisms.
\item We develop a principled security framework for synchronization authentication under real-time and multicast constraints, and design \sys, a scalable and timing-aware protection mechanism.
\item We implement a prototype of \sys and evaluate it in a realistic O-RAN testbed, demonstrating strong integrity and authenticity guarantees while preserving high-precision synchronization performance under adversarial manipulation and packet loss.
\end{itemize}
By systematically closing a previously underexplored vulnerability in O-RAN synchronization, this work provides a practical and principled foundation for securing next-generation open radio access networks.

\section{Background}~\label{sec:bg}
In this section, we first outline the O-RAN architecture and the structure of the Open Fronthaul (O-FH), then describe the synchronization mechanisms used to maintain precise timing across O-RAN deployments, and finally discuss existing protection approaches and their limitations.

\subsection{O-RAN Architecture and Open Fronthaul}

O-RAN adopts a disaggregated architecture that separates traditional base-station functionality into standardized components, enabling multi-vendor interoperability and flexible deployment~\cite{polese2023understanding,oran_3gpp_split}. Among the functional splits defined by the O-RAN Alliance, Split~7-2x is widely adopted: upper-PHY and MAC functions reside at the O-RAN Distributed Unit (O-DU, abbreviated as DU hereafter), while lower-PHY processing is executed at the O-RAN Radio Unit (O-RU, abbreviated as RU hereafter), which interfaces with the radio front end. 
The DU and RU communicate over the Open Fronthaul (O-FH), a packet-based Ethernet interface organized into four logical planes: Control, User, Management, and Synchronization. The Synchronization Plane distributes timing and phase information essential for coordinated baseband and radio operation, forming the temporal foundation of fronthaul.

\begin{figure}[t]
    \centering
    \includegraphics[width=0.9\linewidth]{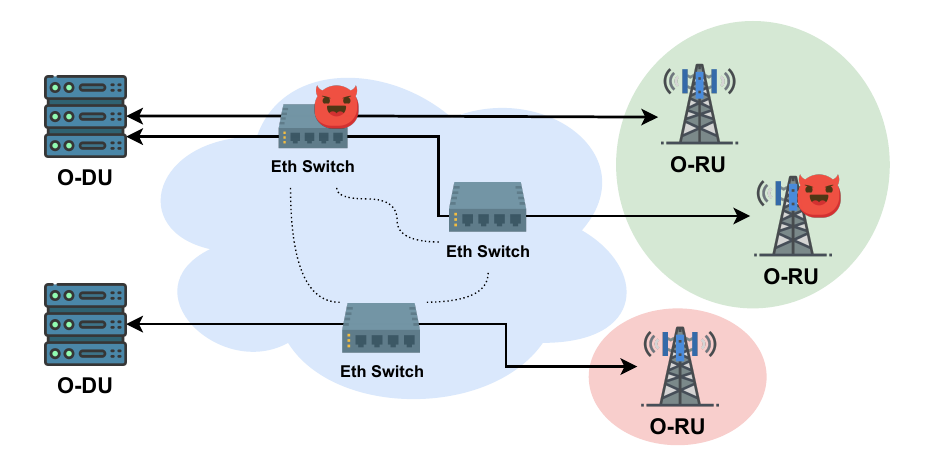}
    \caption{Example O-RAN fronthaul topology and attack surface for PTP synchronization}
    \label{fig:ofh_arch}
\end{figure}

\subsection{Synchronization in O-RAN}

Precise timing in O-RAN is typically achieved using the IEEE 1588v2 Precision Time Protocol (PTP)~\cite{ptp1}. PTP follows a hierarchical master–slave model in which a master clock distributes reference time and slaves synchronize using timestamped message exchanges. From these timestamps, each slave estimates clock offset and path delay relative to the master and disciplines its local oscillator via a feedback controller (clock \emph{servo}).

\noindent\textbf{Delay Request-Response mechanism.} 
O-RAN deployments adopt the Delay Request--Response mechanism, in which two independent message pairs serve distinct roles: \textit{Sync}/\textit{Follow\_Up} carry master-to-slave timing information for offset estimation, while \textit{Delay\_Req}/\textit{Delay\_Resp} measure the slave-specific path delay and may operate at a different rate. Path delay and clock offset are computed as 
$\text{Delay} = \frac{(T_2-T_1)+(T_4-T_3)}{2}$ and  $\text{Offset} = (T_2-T_1) - \text{Delay}$
where $(T_1,T_2)$ are the transmission and reception times of \textit{Sync} ($T_1$ conveyed via \textit{Follow\_Up} in two-step mode), and $(T_3,T_4)$ are the transmission and reception times of \textit{Delay\_Req}. Although PTP also specifies a Peer Delay mechanism for link-local measurements, O-RAN fronthaul synchronization primarily relies on this master--slave model, which we adopt throughout.

\noindent
\textbf{Clock Servo Dynamics.}
Offset and delay estimates are filtered to mitigate timestamp noise and packet delay variation~\cite{linuxptp}. The filtered offset drives a closed-loop servo, typically based on a proportional–integral (PI) controller~\cite{ptp4l}, which computes bounded frequency adjustments to gradually reduce phase error. When offsets exceed configured thresholds, discrete phase steps may be applied; otherwise, the servo operates in frequency-slew mode. Importantly, the control input should be bounded, ensuring stable phase evolution and preventing unbounded timing excursions.

\noindent
\textbf{PTP Roles in O-RAN Deployments.}
As shown in Figure~\ref{fig:ofh_arch}, a transport-network grandmaster provides the primary time reference; the DU acts as PTP master within the fronthaul segment; and RUs act as slaves. Intermediate switches may operate as Boundary or Transparent Clocks, forwarding or compensating PTP traffic depending on deployment configuration.

\subsection{Protection Mechanisms in PTP}

Although PTP can achieve sub-microsecond synchronization accuracy under ideal conditions, its messages are typically transmitted in plaintext without mandatory integrity or authenticity protection. Originally designed for closed and trusted environments, PTP does not inherently defend against adversarial manipulation. Consequently, unauthenticated synchronization traffic is vulnerable to spoofing, replay, and delay manipulation attacks~\cite{finkenzeller2024ptpsec,itkin2017security,moussa2016detection,moussa2018securing}.

\noindent
\textbf{Conventional Security Approaches.}
Several mechanisms aim to enhance PTP message protection.
IEEE~1588 Annex~K defines an optional \textit{AUTHENTICATION} TLV with a sequence counter and a shared-key HMAC, but it relies on pre-shared keys, lacks standardized key management, and scales poorly to multicast deployments.
Network-layer mechanisms such as TLS, IPsec, and MACsec~\cite{mizrahi2011time,dik2023open} provide mature authenticated channels, but they are primarily designed for pairwise protection.
In high-rate fronthaul synchronization, pairwise channels introduce per-destination processing and serialization at the DU, while multicast support is limited or falls back to shared group keys.
These mechanisms also retain a symmetric-key exposure model that is ill-suited to partially trusted RUs.
Any long-lived symmetric key stored at an RU becomes a future-forgery capability once extracted from memory.
A pairwise DU--RU key confines the damage to the compromised association but still enables future authenticated forgeries until rekeying completes; a shared multicast key amplifies the same failure to the entire RU group.
Thus, the limitation is not only overhead or multicast inefficiency, but also the lack of forward compromise containment for receiver-held symmetric secrets.
Lightweight primitives such as ASCON~\cite{dobraunig2021ascon} can reduce cryptographic cost, but do not by themselves address this key-exposure problem.

\noindent
\textbf{TESLA: Lightweight Broadcast Authentication}
TESLA (Timed Efficient Stream Loss-Tolerant Authentication)~\cite{perrig2003tesla,maftei2018implementing,shereen2019next} provides scalable broadcast authentication using delayed key disclosure. The sender constructs a one-way key chain $(K_0, \dots, K_n)$ with $K_i = F(K_{i+1})$, and authenticates messages in interval $i$ using key $K_i$, which is disclosed after a fixed delay. Receivers buffer messages and verify them once the corresponding key is revealed.

TESLA eliminates pairwise key establishment and avoids long-lived shared group secrets, providing scalability and forward security in multicast settings. However, verification latency is inherent to its design: messages must be buffered until key disclosure. In tightly coupled real-time systems such as O-RAN synchronization, this delayed-verification model conflicts with immediate control-loop updates.

\subsection{Challenges in O-RAN}
Applying existing protection mechanisms to O-RAN fronthaul synchronization is challenging due to the combination of multi-RU coordination, partial trust, and strict real-time determinism.

\noindent
\textbf{Multi-RU synchronization under limited trust.}
A single DU typically synchronizes multiple RUs that must remain aligned within a common timing epoch. For features such as TDD and CoMP~\cite{comp}, even small inter-RU skew can cause cross-link interference or reduce coordination gains. Synchronization must therefore provide both high accuracy and deterministically bounded skew across RUs.
Pairwise unicast authentication secures each DU--RU association independently, but it also introduces per-destination processing, queueing, and serialization at the DU, which can amplify arrival-time divergence under load. Multicast dissemination better preserves group simultaneity, but a shared symmetric group key creates compromise amplification: a single compromised RU can forge synchronization traffic for the entire group.

\noindent
\textbf{Compromise containment.}
RUs are distributed edge devices deployed across heterogeneous and potentially partially trusted environments~\cite{groen2024timesafe}; partial endpoint compromise must therefore be treated as a realistic threat. If a receiver stores long-lived symmetric secrets, memory disclosure grants future-forgery capability: group keys compromise all receivers, while pairwise keys compromise the affected DU--RU association until rekeying. Synchronization protection should therefore expose only verification material for expired rounds.


\noindent
\textbf{Strict real-time determinism.}
Synchronization updates directly drive the clock servo at each RU and operate at sub-microsecond granularity. Protection mechanisms that introduce buffering, additional processing, or delayed application risk perturbing the control loop and degrading timing stability. Broadcast authentication schemes such as TESLA provide scalability and avoid receiver-held signing-capable secrets, but conventional verify-then-apply TESLA imposes verification latency. Conversely, pairwise protections can authenticate immediately but increase serialization and processing overhead at the DU.

Taken together, these constraints require a mechanism that simultaneously (i)~supports scalable one-to-many authentication, (ii)~contains RU compromise without relying on receiver-held future-forgery secrets, and (iii)~preserves real-time determinism with bounded inter-RU skew. Among existing approaches, TESLA most naturally addresses (i) and (ii): its broadcast-compatible design avoids pairwise keying, and its one-way key chain ensures that receivers learn only expired verification keys rather than keys for future message authentication. However, conventional TESLA's verify-then-apply semantics fundamentally conflict with (iii), motivating the design of \sys.

\section{Threat Model}\label{sec:threat}

We define a threat model tailored to PTP synchronization in disaggregated O-RAN fronthaul deployments. Our assumptions are grounded in prior empirical analyses of PTP vulnerabilities and timing attacks~\cite{rezabek2022ptp,groen2024timesafe}, as well as documented security challenges in O-RAN systems. We consider realistic adversaries operating in partially trusted fronthaul environments, where synchronization traffic is often unauthenticated due to deployment and performance constraints.

\subsection{Adversary Capabilities}

We consider an adversary targeting the synchronization plane (S-plane) with the objective of degrading timing precision, inducing inter-RU skew, or destabilizing synchronization control loops. The adversary does not control all fronthaul planes but can interfere with PTP traffic traversing the switched fronthaul path.
As illustrated in Figure~\ref{fig:ofh_arch}, we model two categories of capabilities.

\noindent
\textbf{Synchronization-path compromise (MITM).}
By compromising intermediate switches or inserting a malicious device along the fronthaul, the adversary can act as a man-in-the-middle on PTP traffic. In this position, the attacker can eavesdrop, inject, forge, drop, or replay PTP packets, and introduce asymmetric delay or controlled latency to selected messages to bias clock-offset and path-delay estimation. This capability is confined to the synchronization plane and does not imply sustained manipulation of high-rate I/Q payload streams or higher-layer control entities. We do not focus on unbounded selective end-to-end delay or complete synchronization blocking, which constitute availability disruptions and have been addressed by prior detection and resilience mechanisms~\cite{groen2024timesafe,alghamd2020detection,moussa2015detection}.

\noindent
\textbf{Endpoint compromise.}
The adversary may partially compromise an RU and read locally stored cryptographic material, including symmetric keys, key schedules, or authentication state. This capability captures practical attacks such as physical access, local privilege escalation, software vulnerabilities, or read-only exposure on edge-deployed RUs. We do not require the compromised RU to be fully Byzantine: transient key disclosure may be feasible without persistent control of the RU's PTP stack, radio behavior, or outgoing traffic, which would require stronger and noisier capabilities. Nevertheless, memory disclosure alone is sufficient to break conventional symmetric authentication. If the RU stores a shared group key, the attacker can generate valid tags for the entire synchronization group; if it stores a pairwise DU--RU key, the attacker can generate valid tags within that association and impersonate authenticated synchronization messages involving that RU. Accordingly, our goal is to contain RU compromise by ensuring that receiver-side state contains only verification material for expired rounds, rather than long-lived secrets capable of authenticating future DU-originated PTP messages.



\noindent
\textbf{Cryptographic assumptions and scope.}
We assume a computationally bounded adversary that cannot break standard cryptographic primitives without obtaining secret keys. Constructions such as HMAC, ASCON, and secure PRFs are treated as secure. Unrevealed TESLA keys remain unpredictable unless an endpoint is compromised. We assume a minimally trusted initialization phase in which the DU provisions key-chain commitments and configuration parameters to each RU. We do not assume confidentiality of PTP traffic; our focus is on scalable, low-overhead guarantees of \emph{authenticity} and \emph{integrity} against spoofing and replay, as well as robustness against adversarial delay manipulation and packet dropping, whose impact is bounded within the authentication window.

\subsection{Practicality and Scope}

O-RAN deployments often span heterogeneous or partially trusted infrastructures, where fronthaul segments may be exposed to network-level manipulation or endpoint compromise. Prior work demonstrates that replay or delay manipulation of PTP traffic can destabilize production-grade 5G systems within seconds by biasing timing loops and misaligning coordinated RUs~\cite{groen2024timesafe}. Notably, these attacks operate solely on low-bandwidth synchronization traffic yet induce large-scale desynchronization. Other studies show that broader fronthaul access may enable manipulation of I/Q samples or control-plane signaling~\cite{xing2024criticality}; however, such attacks require substantially stronger capabilities and target different planes.
Endpoint compromise further expands the attack surface. Physical or local network access to O-RAN RUs may expose configuration data and cryptographic material~\cite{janzen2024oh}. Mechanisms relying on shared group keys are particularly vulnerable to compromise amplification, allowing a single breached RU to affect all receivers.

Motivated by these findings, our threat model captures both synchronization-path manipulation and partial endpoint compromise under realistic deployment conditions. We do not assume fully trusted endpoints and explicitly account for compromise containment. Our goal is to close a documented integrity gap in the synchronization plane that alone can induce large-scale service disruption. This work complements, rather than replaces, protections for user and control planes.

\section{Detailed Design}~\label{sec:design}

Our goal is to provide strong \emph{integrity} and \emph{authenticity} guarantees for PTP synchronization while preserving O-RAN's stringent real-time constraints. The design is guided by three principles: (i)~minimizing trust in intermediaries and endpoints; (ii)~preserving deterministic multi-RU timing alignment; and (iii)~avoiding heavyweight cryptography that perturbs synchronization precision, while ensuring graceful degradation under partial compromise.

We build on TESLA because delayed disclosure removes future-forgery secrets from receivers while supporting multicast authentication. Conventional TESLA, however, buffers messages until key disclosure, which delays servo inputs and destabilizes PTP control. PRTESLA-C instead applies samples immediately, verifies them after disclosure, and removes invalid samples from persistent synchronization state through bounded recovery.

We organize the design around three dimensions: key-disclosure granularity, synchronization computation, and authentication timing. These choices lead to \textbf{Per-Round TESLA with Correction (\sys)}, a delayed-authentication mechanism tailored to O-RAN PTP synchronization.

\subsection{Design Space}
\label{sec:design-space}

We characterize the design space along three dimensions that capture the main trade-offs in securing PTP synchronization over O-RAN fronthaul.

\smallskip
\noindent\textbf{Dimension 1: Key-Disclosure Granularity.}
This dimension determines how authentication keys are assigned, rotated, and disclosed, and therefore how widely a compromise can propagate:
\emph{(a) Shared keys} are efficient but amplify RU compromise. \emph{(b) Per-interval keys} reduce management cost but introduce boundary ambiguity under loss and reordering. \emph{(c) Per-round keys} avoid interval-boundary ambiguity and bound failures to individual rounds; PRTESLA-C adopts this choice with separate Sync and Delay chains.

\smallskip
\noindent\textbf{Dimension~2: Synchronization Computation.}
This dimension determines how timing estimates are computed when multiple rounds elapse before key disclosure. Per-interval schemes may use \emph{(a) last-only} samples, where only the last valid round is used, minimizing buffering but increasing sensitivity to loss, jitter, and interval-boundary artifacts; or \emph{(b) aggregated} samples, where multiple rounds are combined, improving smoothing but adding buffering and recovery complexity. In \sys, per-round authentication avoids authentication-induced interval aggregation: each completed PTP exchange contributes independently.




\smallskip
\noindent\textbf{Dimension~3: Authentication Application Timing.}
This dimension determines \emph{when} a sample affects the local clock. In \emph{(a) verify-then-apply}, the receiver waits for key disclosure and MAC verification before applying the sample, avoiding unauthenticated servo inputs but introducing stale control inputs. In \emph{(b) apply-then-verify}, the receiver applies the sample immediately, verifies it after disclosure, and corrects persistent state on failure. This creates a short speculative window, analyzed in Section~\ref{sec:timing-analysis}, but preserves PTP servo timing; \sys adopts this model.




\begin{figure}[t]
    \centering
    \includegraphics[width=0.9\linewidth]{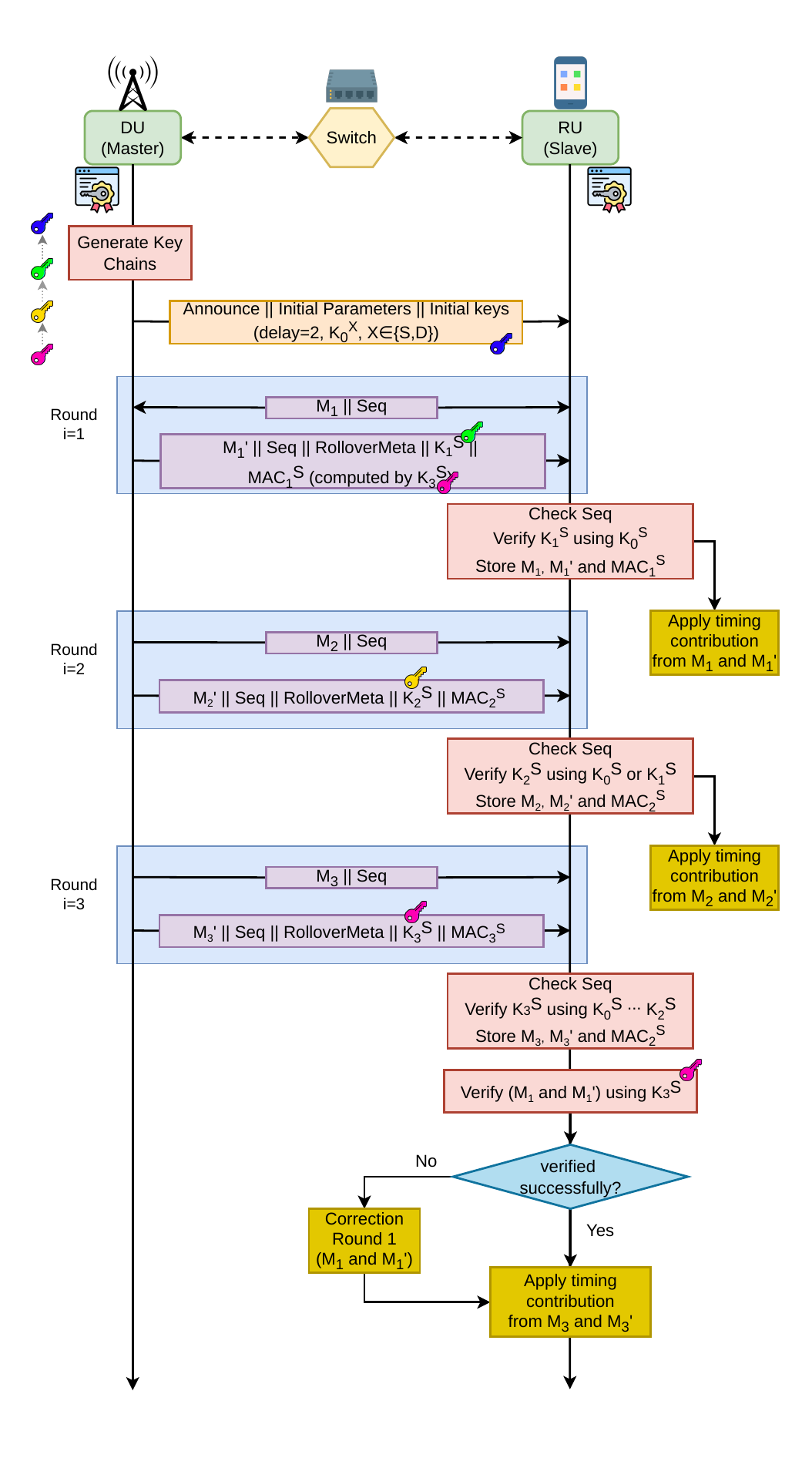}
    \caption{\sys Workflow.
    The same workflow applies independently to the Sync domain (S) and Delay domain (D). $M_i^S$ binds Sync/Follow\_Up fields, while $M_i^D$ binds Delay\_Req/Delay\_Resp fields and the requesting RU identity.
    }
    \label{fig:workflow}
\end{figure}

\subsection{\sys: Per-Round TESLA with Correction}
\label{sec:sys}

Guided by the design taxonomy in Section~\ref{sec:design}, we present \sys (\textbf{Per-Round TESLA with Correction}), a per-round delayed-authentication protocol for securing PTP synchronization in O-RAN. \sys integrates (i) \emph{per-round key disclosure} aligned with the PTP four-message exchange, 
(ii)~\emph{independent
per-domain authentication} of Sync and Delay contributions each
round,
and (iii) an \emph{apply-then-verify-and-correct} execution model that preserves real-time responsiveness while enabling retrospective integrity enforcement.
We describe \sys as a sequence of phases executed by the DU (PTP master) and each RU (PTP slave). 

\subsubsection{Phase~1: Setup and Bootstrapping}
\label{sec:sys-setup}

The DU initializes two independent one-way key chains, one per authentication domain: the \emph{Sync domain} ($\mathcal{D}_S$), covering \textit{Sync} and \textit{Follow\_Up} messages, and the \emph{Delay domain} ($\mathcal{D}_D$), covering \textit{Delay\_Req} and \textit{Delay\_Resp} exchanges. 
Each chain $\{K_N^{(\cdot)},\allowbreak K_{N-1}^{(\cdot)},\allowbreak \ldots,K_0^{(\cdot)}\}$ is constructed via a cryptographic one-way function $F$ with $K_i = F(K_{i+1})$, with domain separation enforced by distinct derivation contexts. The chain anchors $K_0^{(S)}$ and $K_0^{(D)}$ are provisioned to each RU over a minimally trusted bootstrap channel, together with the disclosure delay~$d$ and an initial round index~$i_0$.
A separate derivation function $F' $ is used to derive per-round MAC keys as $K'_i = F'(K_i)$, ensuring that the MAC key and the chain key are domain-separated and that knowledge of a MAC key does not compromise the key-chain structure.


\subsubsection{Phase~2: Round Execution and Immediate Application}
\label{sec:sys-apply}


In steady state, \sys authenticates each round as two independent domain contributions: the Sync domain covering \textit{Sync}/\textit{Follow\_Up}, and the Delay domain covering \textit{Delay\_Req}/\textit{Delay\_Resp}. The two pairs serve distinct roles and may operate at different exchange rates: \textit{Sync}/\textit{Follow\_Up} carry master-to-slave timing information for offset estimation, while \textit{Delay\_Req}/\textit{Delay\_Resp} perform a slave-initiated round-trip for path-delay measurement. A unified bundle would couple their buffering and disclosure behavior, causing out-of-order or rate-mismatched arrivals to break bundle assembly. Separate domains prevent failures or replays in one direction from contaminating the other's authentication state, and scope recovery accordingly: Sync failures reset servo state, while Delay failures rebuild the path-delay estimator.



\smallskip
\noindent\textbf{Sync domain.}
Let $M_i^{(S)}$ denote the canonical payload constructed from the timing-sensitive fields of the \textit{Sync} and \textit{Follow\_Up} messages of round~$i$ (timestamps, correction fields, source port identity, and sequence identifier). After transmitting \textit{Follow\_Up}, the DU computes
\[
  \mathrm{MAC}_i^{(S)} = \mathrm{ASCON}\!\left(F'(K_i^{(S)}),\; M_i^{(S)}\right)
\]
and appends $\mathrm{MAC}_i^{(S)} \Vert K_{i-d}^{(S)} \Vert \textit{RolloverMeta}_i^{(S)}$ to \textit{Follow\_Up}, leaving the high-rate \textit{Sync} message unmodified to minimize overhead on
the most latency-sensitive message type. $\textit{RolloverMeta}_i^{(S)}$ carries the next-epoch chain commitment and switch point for the Sync domain, and is non-empty only when the current epoch is nearing exhaustion (Section~\ref{sec:sys-rollover}).

\smallskip
\noindent\textbf{Delay domain.}
Let $M_i^{(D)}$ denote the canonical payload constructed from the \textit{Delay\_Req} and \textit{Delay\_Resp} messages of round~$i$. To prevent cross-slave replay, $M_i^{(D)}$ explicitly binds the requesting RU's port identity $\mathrm{PID}_{\mathrm{RU}}$, ensuring that a \textit{Delay\_Resp} authenticated for one slave cannot be accepted by another. After completing the exchange, the DU computes
\[
  \mathrm{MAC}_i^{(D)} = \mathrm{ASCON}\!\left(F'(K_i^{(D)}),\; M_i^{(D)}\right)
\]
and appends $\mathrm{MAC}_i^{(D)} \Vert K_{i-d}^{(D)} \Vert \textit{RolloverMeta}_i^{(D)}$ to \textit{Delay\_Resp}, leaving \textit{Delay\_Req} unmodified. Symmetrically, $\textit{RolloverMeta}_i^{(D)}$ carries the next-epoch commitment and switch point for the Delay domain independently, since the two domains may exhaust their respective key chains at different times when operating at different exchange rates.

\smallskip
\noindent\textbf{Fast path and authentication state.} Upon receiving an augmented \textit{Follow\_Up} or \textit{Delay\_Resp}, the RU records the authentication material in the corresponding TESLA stream. Each stream maintains independent state comprising the domain, epoch, sequence identifier, canonical payload, MAC tag, and disclosed-key information; the sequence identifier associates the two messages of each authenticated pair, while the domain and epoch metadata ensure correct verification context.

Crucially, delayed authentication does not stall the PTP pipeline. Samples that pass PTP matching and metadata checks are applied immediately: the RU derives offset and path-delay estimates from the received timestamps and updates the clock servo exactly as in standard PTP. The associated authentication and recovery context is recorded in a ledger entry~$\mathcal{L}_i$; for Delay samples, $\mathcal{L}_i$ additionally retains the state needed to reconstruct verified delay history. Once the corresponding key is disclosed, the buffered materials in~$\mathcal{L}_i$ are verified independently per domain; failure triggers the recovery procedure described in Section~\ref{sec:sys-correct}.
To support multi-RU deployments, all RUs process every \textit{Delay\_Resp} to advance shared authentication state, but apply path-delay updates only from responses whose \texttt{requestingPortIdentity} matches their own port.

\subsubsection{Phase~3: Delayed Verification}
\label{sec:sys-verify}

When the RU receives a disclosed key $K_{i-d}^{(\cdot)}$ piggybacked on a later round, it validates the key incrementally against the last accepted disclosure $K_{\text{last}}^{(\cdot)}$:
\[
  F^{\Delta}(K_{i-d}^{(\cdot)}) = K_{\text{last}}^{(\cdot)}
\]
where $\Delta$ is the index gap between the two disclosures.
This check is $O(\Delta)$, reducing to a single $F$ evaluation under normal operation ($\Delta=1$), and is performed independently per domain. If valid, the RU derives the MAC key $K_{i-d}^{\prime(\cdot)} = F'(K_{i-d}^{(\cdot)})$, recomputes the MAC over the buffered canonical payload, and compares it against the stored tag. Successful verification marks the domain contribution as authenticated; the Sync and Delay domains are verified and released independently.

Loss of a disclosed key or a bundle component causes the affected sample to time out at deadline $W$ and be treated as a verification failure, triggering recovery (Section~\ref{sec:sys-correct}). Crucially, a missing disclosure does not invalidate subsequent rounds: since the incremental check uses $K_{\text{last}}^{(\cdot)}$ rather than a fixed anchor, the receiver skips the gap and resumes verification from the next successfully received key.

\subsubsection{Phase~4: Correction upon Verification Failure}
\label{sec:sys-correct}

If delayed verification fails in either domain, due to a MAC mismatch, invalid key disclosure, missing bundle components, or a verification timeout, \sys applies a control-level recovery action to prevent unauthenticated material from continuing to bias the RU clock. The ledger entry~$\mathcal{L}_i$ retains the authentication context for the affected timing sample and, for Delay-domain samples, the state needed to reconstruct verified delay history. \sys does not roll back wall-clock time, as doing so would disrupt higher-layer operation; instead, recovery removes unauthenticated samples from persistent estimator and servo state, preventing their influence from accumulating beyond the disclosure window.

Recovery is scoped to the affected domain. For \textbf{Delay-domain failures}, \sys resets the path-delay estimation state and reconstructs it by replaying previously verified Delay samples in order, excluding the rejected sample while preserving authenticated history. For \textbf{Sync-domain failures}, \sys resets the affected clock-servo and timestamp-processing state without replaying prior Sync samples, since replaying stale offsets into a closed-loop servo may reintroduce phase disturbances. Subsequent synchronization resumes from fresh samples, and any physical phase deviation induced during the speculative window is treated as transient and corrected by the servo over time. This coarse-grained, domain-scoped recovery removes the persistent influence of invalid material while keeping the time-critical PTP processing path non-blocking.

\subsubsection{Phase~5: Key-Chain Rollover}
\label{sec:sys-rollover}

\sys organizes each domain's key chain into fixed-length epochs, with rollover proceeding independently per domain, as the Sync and Delay chains may be exhausted at different times when operating at different exchange rates. Because key disclosure lags transmission by $d$ rounds, the final $d$ samples of an epoch are verified only after the epoch boundary. \sys handles this via a \emph{carry-over disclosure} mechanism: the first $d$ TLVs of the new epoch piggyback the previous-epoch disclosed keys needed to verify those tail samples, closing the verification gap at the boundary.

$R$ rounds before exhaustion, the DU preannounces the next chain's public commitment via $\textit{RolloverMeta}^{(\cdot)}$ (Section~\ref{sec:sys-apply}). The receiver caches this commitment and accepts next-epoch keys only after validating them against it, guarding against epoch-boundary forgery. The transmitter advances the epoch once the current chain is exhausted; the receiver follows upon successful validation. This design rotates key chains per domain without interrupting the PTP pipeline or altering~$d$.

\subsection{Timing Analysis}
\label{sec:timing-analysis}

\noindent\textbf{Speculative window.}
Let $f_S$ and $f_D$ denote the exchange rates of the Sync and Delay domains. With disclosure delay $d$, authentication material in domain $X \in \{S,D\}$ becomes verifiable after $d$ domain samples, yielding a speculative window
\[
  W_X \approx \frac{d}{f_X} + \tau_X
\]
where $\tau_X$ captures bounded propagation, scheduling, and processing delay. At most $d$ samples per domain remain speculative at any time. For system-level bounds, we take the worst case across domains, $W = \max(W_S, W_D)$.

\noindent\textbf{Bounded transient deviation.}
Let $x(t)$ denote the RU phase error, and let the clock servo enforce a bounded control input $|u(t)| \leq S_{\max}$. Any unauthenticated influence accumulated during the speculative window is bounded by
\[
  |\Delta x| \leq S_{\max} W, \qquad W = \max(W_S, W_D)
\]
Upon verification failure, \sys applies the control-level recovery of Section~\ref{sec:sys-correct}, ensuring that rejected samples cannot persistently bias synchronization state beyond this bound.

\subsection{Security Analysis}
\label{sec:security}

\sys provides scalable one-to-many authentication for PTP synchronization, bounding the window during which unauthenticated material may influence clock state to $W$. Confidentiality is out of scope and can be layered independently, e.g., via link-layer protection.

\noindent\textbf{Authenticity and integrity.}
Each domain contribution is authenticated under key $K_i^{(\cdot)}$ and verified upon delayed disclosure. Under standard assumptions on the MAC and one-way function, an adversary without the undisclosed key cannot forge a valid tag; tampering is therefore detected at verification time. On failure, \sys applies domain-scoped recovery:
Sync failures reset servo and timestamp-processing state, while Delay failures rebuild the path-delay estimator from verified history. Forged or modified samples thus have at most bounded transient influence and cannot persistently bias synchronization state.

\noindent\textbf{Replay and reordering.}
Each authenticated bundle is scoped by domain, epoch, and sequence identifier, ensuring buffered material is verified in the correct context. Replayed or reordered bundles inconsistent with this scope are rejected before or upon delayed verification; any transient influence is bounded by $W$ and removed by the recovery procedure.
Carry-over disclosure at epoch boundaries preserves verification coverage for the final $d$ tail samples, eliminating gaps during key-chain rollover.

\noindent\textbf{Delay manipulation.}
An on-path adversary may bias offset or path-delay estimation by injecting asymmetric or selective delay without modifying packet contents. Payload modifications are detected as integrity failures at verification time. Schedule-consistent delay on otherwise authentic packets is not eliminated by authentication alone: such packets may transiently bias the servo or delay estimator within $W$, with phase deviation bounded by $|\Delta x| \leq S_{\max}W$ (Section~\ref{sec:timing-analysis}). Manipulations that prevent bundle completion or key disclosure before the verification deadline are treated as availability disruptions and trigger recovery.

Against a persistent adversary that repeats this across consecutive windows, \sys's recovery procedure (Section~\ref{sec:sys-correct}) resets the affected servo or delay-estimator state at each verification boundary, preventing bias from accumulating across windows. The long-term effect is therefore a degraded, but bounded, accuracy within each window, not unbounded drift. This residual exposure is a fundamental limitation of delayed authentication shared by all TESLA-based schemes; stronger guarantees require orthogonal mechanisms such as delay-asymmetry consistency checks~\cite{groen2024timesafe,alghamd2020detection,moussa2015detection}.

\noindent\textbf{Packet loss and dropping.}
Packet loss reduces the rate of valid timing samples but does not introduce persistent authenticated bias. Missing bundle components or undisclosed keys cause the affected sample to time out and be treated as a verification failure; the deadline $W$ ensures speculative state is resolved within a bounded interval, and the recovery procedure prevents incomplete samples from continuing to influence RU state.


\noindent\textbf{Forward security and compromise containment.}
\sys provides compromise containment by ensuring that RUs do not hold
long-lived secrets capable of authenticating future DU-originated
synchronization messages. Each authentication domain uses a one-way key
chain, and disclosing $K_{i-d}$ does not allow an adversary to recover any
undisclosed key $K_j$ for $j > i-d$, assuming $F$ is one-way. Therefore,
keys learned by receivers are useful only for verifying expired rounds, not
for generating valid tags for future rounds. Even if an adversary partially
compromises an RU and extracts its local authentication state, future
rounds remain unforgeable except within the bounded speculative window $W$.

This differs fundamentally from conventional symmetric-key protection.
With a shared group MAC key, every RU stores the same long-lived
signing-capable secret; compromising a single RU therefore enables the
adversary to forge synchronization traffic for the entire multicast group
until the group is rekeyed. Pairwise symmetric MACs reduce this blast radius
but do not eliminate the underlying exposure. If the adversary extracts the
DU--RU key from a compromised RU, that key remains usable to generate
valid authentication tags for future synchronization messages within the
compromised association until rekeying completes. Thus, pairwise keying
confines compromise to one DU--RU association, but still grants a persistent
future-forgery capability for that association.

In contrast, \sys changes the receiver-side exposure model. RUs learn
only delayed disclosure keys after the corresponding rounds have expired for
transmission, and the one-way key chain prevents deriving future keys from
disclosed ones. Consequently, RU memory disclosure exposes at most expired
verification material and bounded speculative state, rather than a persistent
secret for future forgery. This per-round disclosure structure directly
limits compromise amplification: a compromised RU cannot forge
synchronization traffic beyond the current disclosure window $W$, and
key-chain rollover further bounds the lifetime of any exposed state to the
current epoch of the affected domain.

\section{Implementation}\label{sec:impl}
We implement PRTESLA-C by extending linuxptp with about 7K lines of C/C++ code. We augment the Sync, Follow Up, Delay Req, and Delay Resp state machines with custom Annex-K TLVs carrying authentication tags, disclosed keys, and rollover metadata. Security logic runs in user space, while timestamping and packet scheduling remain on the kernel fast path, preserving compatibility with hardware timestamping. Key-chain evolution, disclosure handling, and verification execute asynchronously to avoid perturbing deterministic timing. We instantiate $F$ and $F'$ with domain-separated ASCON-based primitives at 128-bit security. Runtime parameters, including $T_{int}$ and disclosure delay $d$, are configurable; our evaluation uses $d=2$. 

\section{Evaluation}
\label{sec:eval}

\subsection{Experimental Setup}
\label{sec:setup}

We evaluate \sys on an O-RAN-like fronthaul testbed interconnected
by an Arista 7010T-48 top-of-rack switch (Figure~\ref{fig:testbed}).
The baseline topology consists of an DU (PTP master), an RU
(PTP slave), and an impairment node for controlled MITM attacks
(replay, loss, and delay manipulation).

\noindent\textbf{Single-RU topology.}
The DU and RU run on dedicated Linux hosts with an Intel
Core i7-3770 CPU, 16\,GB RAM, and a Mellanox ConnectX-4 Lx EN
NIC (MCX4121A-ACAT, $2{\times}25$\,GbE), providing hardware
PTP/PHC timestamping via \texttt{SO\_TIMESTAMPING}. All nodes
run Ubuntu~22.04 (kernel~6.8) with \texttt{linuxptp}~v4.4~\cite{linuxptp}
over 25\,GbE full-duplex links. This topology is used for all
baseline comparisons and attack experiments.

\noindent\textbf{Multi-RU topology.}
To study one-to-many synchronization, we extend the setup to a
single-master, two-slave topology within the same L2 broadcast
domain. Two logical RUs are emulated via isolated Linux network
namespaces on a shared host, connected to the master-facing network
through bridged virtual interfaces to preserve multicast PTP
dissemination. The DU retains hardware timestamping; the
namespace-based RUs use software timestamping, approximating
two independent RUs on the same switch.

\noindent\textbf{End-to-end compatibility.}
To confirm that \sys does not interfere with higher-layer O-RAN
components, we perform E2E functional validation using an
srsRAN-based pipeline: a DU-side gNB (\texttt{gnb-oran}) and an
RU emulator (\texttt{ru\_emulator}) interconnected via a Linux
\texttt{veth} pair, with \sys enabled on the synchronization path.
Validation succeeds if (i)~the RU emulator initializes and binds
to the fronthaul interface, (ii)~the gNB completes OFH setup, and
(iii)~the system runs continuously without synchronization loss,
RU--DU interface errors, or abnormal PTP behavior.

\begin{figure}[t]
    \centering
    \includegraphics[width=0.85\linewidth]{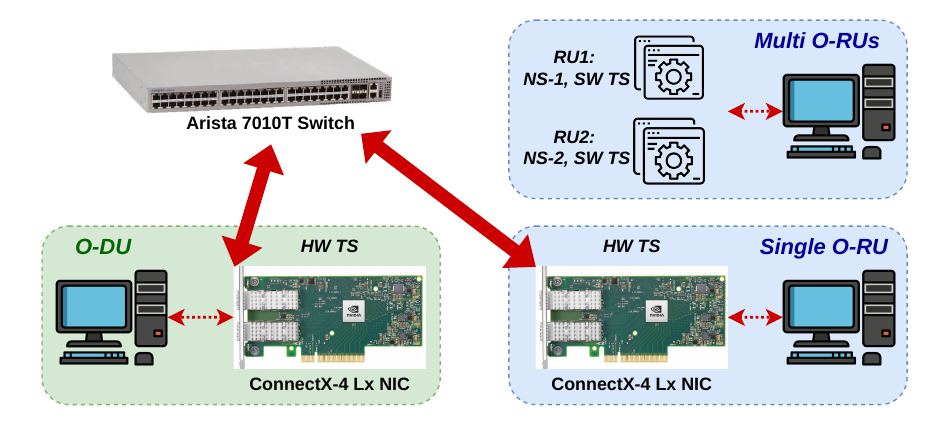}
    \caption{O-RAN PTP evaluation testbed: DU (master) and RU (slave) interconnected via a switch.
    }
    \label{fig:testbed}
\end{figure}

\subsection{Baselines and Metrics}
\label{sec:baselines-metrics}

We evaluate \sys along four axes: \emph{timing accuracy},
\emph{timing stability}, \emph{runtime overhead}, and
\emph{resilience to active attacks}. All experiments use identical
hardware, background traffic, and synchronization parameters.
Each experiment is repeated 5 times; we report means with
variability where appropriate.

\noindent\textbf{Baselines.}
We compare against \textit{S0}~(plain IEEE~1588, no authentication); \textit{S1}~(Annex~K-style shared-key authentication with AES-GCM (\textbf{S1-1}, as in MACsec~\cite{dik2023open}), HMAC-SHA256 (\textbf{S1-2}, native to \texttt{linuxptp}), and ASCON-MAC (\textbf{S1-3})); \textit{S2}~(TESLA-style delayed disclosure with ASCON under \emph{verify-then-apply} semantics, at three keying granularities: per-interval last-round (\textbf{S2-1}), per-interval averaged (\textbf{S2-2}), and per-round (\textbf{S2-3})); and \textit{\sys}~(per-round disclosure with \emph{apply-then-verify-and-correct} and ledger-based correction).
Because S2 variants require per-scheme instrumentation and timestamping assumptions incompatible with namespace-based emulation, the full baseline comparison is restricted to the single-RU topology. Multi-RU experiments compare S1-1 (MACsec-compatible) and S1-3 (standard ASCON-based) as representative shared-key baselines alongside \sys. 
E2E experiments compare \sys against S0 only, focusing on functional correctness and timing preservation under a realistic O-RAN stack.

\noindent
\textbf{Metrics.}
We evaluate timing precision and control-loop behavior via three primary metrics: \emph{RMS clock offset}, capturing synchronization accuracy (mean offset vanishes in steady state, making RMS the appropriate measure of residual deviation); \emph{RMS frequency correction}, reflecting servo stability (large RMS indicates control-loop oscillation); and \emph{mean one-way delay}, assessing path-delay estimation quality. Standard deviations are reported alongside each metric to quantify run-to-run variability. In the single-RU topology, we additionally measure CPU and memory usage, packet-size overhead, and microarchitectural indicators (IPC, stall cycles, context switches), and assess resilience under active attacks.

\subsection{Single-RU Evaluation}
\label{sec:single-ru}

\subsubsection{Timing Performance: Accuracy and Stability}
\label{sec:single-ru-timing}

Figure~\ref{fig:timeacc} reports RMS clock offset, frequency correction, and one-way delay across all schemes.
Shared-key schemes (S1-x) match S0 closely across all metrics, as symmetric MAC verification completes off the critical path without delaying timestamp acquisition or servo updates.

All S2-x variants exhibit catastrophic timing degradation, that is, RMS offsets and frequency corrections exceed S0 by six orders of magnitude. The root cause is structural: withholding servo updates until key disclosure forces the control loop to operate on stale timing, introducing an effective latency $W \approx d/r + \tau$ that far exceeds the servo's stability margin. Notably, S2-3 (per-round TESLA) offers no improvement over per-interval variants; under \emph{verify-then-apply} semantics, finer disclosure granularity stalls the servo more frequently, amplifying jitter rather than reducing it. This underscores that the decisive factor is not disclosure granularity but whether servo updates are decoupled from the authentication path.

\sys achieves timing accuracy and stability indistinguishable from S0 and S1-x across all metrics. Synchronization updates are applied immediately at the native round cadence; authentication executes asynchronously without touching the control-loop path, and the per-round correction mechanism absorbs rejected rounds without disrupting servo operation. Together, these results confirm that apply-then-verify-and-correct successfully decouples authentication latency from synchronization responsiveness.

\begin{figure*}[t]
  \centering
  \includegraphics[width=1.0\linewidth]{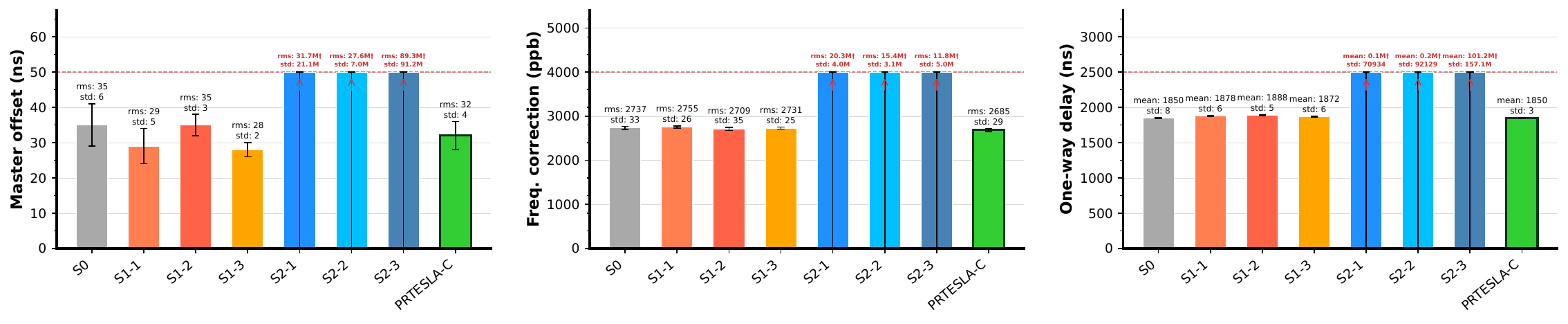}
  \caption{Timing accuracy across 
  authentication schemes.}
  \label{fig:timeacc}
\end{figure*}

\subsubsection{Overhead: Runtime, Bandwidth, and Microarchitectural Effects}
\label{sec:single-ru-overhead}

\noindent\textbf{Bandwidth overhead.}
Table~\ref{tab:ptp-packets} shows per-message TLV sizes across schemes. Shared-key schemes (S1-x) attach a MAC to every message uniformly, imposing overhead on the high-rate \textit{Sync}, \textit{Delay\_Req} path regardless of message frequency. \sys, like S2, concentrates authentication metadata in lower-rate messages: \textit{Sync} and \textit{Delay\_Req} carry only a compact sequence tag, while MACs, disclosed keys, and rollover metadata are confined to \textit{Follow\_Up}, \textit{Delay\_Resp}, and \textit{Announce}. This asymmetry is deliberate, that is, fronthaul bandwidth is most sensitive to overhead on the highest-frequency messages, and \sys preserves that path with minimal augmentation.

\noindent\textbf{Runtime overhead.}
As shown in Table~\ref{tab:overhead}, all schemes consume well under 1\% CPU on both DU and RU, confirming that authentication is computationally lightweight in practice. On the DU, differences across schemes are negligible, as the sender's workload is dominated by key lookup and tag generation rather than verification. On the RU, \sys incurs modest CPU usage comparable to shared-key schemes, which is a natural consequence of combining immediate application with asynchronous per-round verification. Memory consumption on the RU follows verification window length: per-interval TESLA variants (S2-1, S2-2) require larger buffers to hold unverified rounds over longer windows, while \sys and S2-3 (per-round) share a substantially smaller footprint, consistent with their bounded, short verification windows.

\noindent\textbf{Microarchitectural efficiency.}
S2-1 and S2-2 exhibit elevated context-switch rates relative to all other schemes, reflecting the scheduling pressure of long buffering windows that require periodic background wake-ups to resolve pending rounds. \sys, by contrast, maintains context-switch rates comparable to the unsecured baseline, as its short, fixed verification window resolves predictably within each round cadence, avoiding irregular wake-up patterns.

\begin{table}[t]
\small
    \centering
    \caption{
        TLV sizes (bytes) for PTP messages across schemes.
    }
    \label{tab:ptp-packets}
    \begin{tabular}{lcccccccc}
        \toprule
        \textbf{Packet} & \textbf{S0} & \textbf{S1-1} & \textbf{S1-2} & \textbf{S1-3} & \textbf{S2} & \textbf{\sys} \\
        \midrule
        Announce   & 0 & 26 & 42 & 26 & 78 & 78 \\
        Sync       & 0 & 26  & 42 & 26 & 10   & 10   \\
        Follow\_Up & 0 & 26  & 42 & 26 & 46   & 46   \\
        Delay\_Req & 0 & 26  & 42 & 26 & 10   & 10   \\
        Delay\_Resp& 0 & 26  & 42 & 26 & 46   & 46 \\
        \bottomrule
    \end{tabular}
\end{table}

%

\newcommand{\minCPUDU}{0.40}  \newcommand{\maxCPUDU}{0.55}
\newcommand{\minCPURU}{0.33}  \newcommand{\maxCPURU}{0.75}
\newcommand{\minMem}{2688}    \newcommand{\maxMem}{6400}
\newcommand{\minIPC}{0.27}    \newcommand{\maxIPC}{0.64}
\newcommand{\minStall}{75.18} \newcommand{\maxStall}{85.92}
\newcommand{\minCtx}{10.55}   \newcommand{\maxCtx}{16.71}

\begin{table}[t]
  \centering
  \caption{Runtime overhead and microarchitectural efficiency across schemes.}
  \label{tab:overhead}
  \setlength{\tabcolsep}{4pt}
  \resizebox{\columnwidth}{!}{%
  \begin{tabular}{l cccccc}
    \toprule
    \multirow{3}{*}{\textbf{Scheme}}
      & \multicolumn{3}{c}{\textbf{Runtime Overhead}}
      & \multicolumn{3}{c}{\textbf{Microarch. Efficiency}} \\
    \cmidrule(lr){2-4}\cmidrule(lr){5-7}
      & {$\text{CPU}_\text{DU}$}
      & {$\text{CPU}_\text{RU}$}
      & {$\text{Mem}_\text{RU}$}
      & {IPC}
      & {Stall}
      & {Ctx-sw} \\
      & {(\%) $\downarrow$}
      & {(\%) $\downarrow$}
      & {(KB) $\downarrow$}
      & {$\uparrow$}
      & {(\%) $\downarrow$}
      & {(K/s) $\downarrow$} \\
    \midrule
    S0
      & \cellLow{0.43}{\minCPUDU}{\maxCPUDU}
      & \cellLow{0.33}{\minCPURU}{\maxCPURU}
      & \cellLow{3968}{\minMem}{\maxMem}
      & \cellHigh{0.35}{\minIPC}{\maxIPC}
      & \cellLow{81.51}{\minStall}{\maxStall}
      & \cellLow{16.71}{\minCtx}{\maxCtx} \\
    S1-1
      & \cellLow{0.40}{\minCPUDU}{\maxCPUDU}
      & \cellLow{0.70}{\minCPURU}{\maxCPURU}
      & \cellLow{6400}{\minMem}{\maxMem}
      & \cellHigh{0.51}{\minIPC}{\maxIPC}
      & \cellLow{78.71}{\minStall}{\maxStall}
      & \cellLow{11.21}{\minCtx}{\maxCtx} \\
    S1-2
      & \cellLow{0.55}{\minCPUDU}{\maxCPUDU}
      & \cellLow{0.75}{\minCPURU}{\maxCPURU}
      & \cellLow{6016}{\minMem}{\maxMem}
      & \cellHigh{0.64}{\minIPC}{\maxIPC}
      & \cellLow{75.18}{\minStall}{\maxStall}
      & \cellLow{10.65}{\minCtx}{\maxCtx} \\
    S1-3
      & \cellLow{0.50}{\minCPUDU}{\maxCPUDU}
      & \cellLow{0.50}{\minCPURU}{\maxCPURU}
      & \cellLow{2688}{\minMem}{\maxMem}
      & \cellHigh{0.27}{\minIPC}{\maxIPC}
      & \cellLow{85.92}{\minStall}{\maxStall}
      & \cellLow{12.72}{\minCtx}{\maxCtx} \\
    S2-1
      & \cellLow{0.45}{\minCPUDU}{\maxCPUDU}
      & \cellLow{0.50}{\minCPURU}{\maxCPURU}
      & \cellLow{6144}{\minMem}{\maxMem}
      & \cellHigh{0.49}{\minIPC}{\maxIPC}
      & \cellLow{77.63}{\minStall}{\maxStall}
      & \cellLow{15.37}{\minCtx}{\maxCtx} \\
    S2-2
      & \cellLow{0.45}{\minCPUDU}{\maxCPUDU}
      & \cellLow{0.55}{\minCPURU}{\maxCPURU}
      & \cellLow{5632}{\minMem}{\maxMem}
      & \cellHigh{0.48}{\minIPC}{\maxIPC}
      & \cellLow{78.15}{\minStall}{\maxStall}
      & \cellLow{15.14}{\minCtx}{\maxCtx} \\
    S2-3
      & \cellLow{0.50}{\minCPUDU}{\maxCPUDU}
      & \cellLow{0.54}{\minCPURU}{\maxCPURU}
      & \cellLow{4224}{\minMem}{\maxMem}
      & \cellHigh{0.38}{\minIPC}{\maxIPC}
      & \cellLow{82.37}{\minStall}{\maxStall}
      & \cellLow{12.56}{\minCtx}{\maxCtx} \\
    \midrule
    \textbf{PRTESLA-C}
      & \cellLow{0.45}{\minCPUDU}{\maxCPUDU}
      & \cellLow{0.73}{\minCPURU}{\maxCPURU}
      & \cellLow{4224}{\minMem}{\maxMem}
      & \cellHigh{0.31}{\minIPC}{\maxIPC}
      & \cellLow{84.82}{\minStall}{\maxStall}
      & \cellLow{10.55}{\minCtx}{\maxCtx} \\
    \bottomrule
  \end{tabular}%
  }
  \vspace{2pt}
  \begin{minipage}{\linewidth}
    \footnotesize
    CPU: \%usr\,+\,\%sys.
    IPC: instructions per cycle.
    Stall: front-end stalled cycles.
    Ctx-sw: context switches/s; elevated in S2-1/S2-2
    due to deferred-verification buffering.
    Means over 5 runs. \\
    $\downarrow$~lower is better, $\uparrow$~higher is better. \\
    Cell color: \textcolor[rgb]{0.78,0.98,0.78}{\rule{6pt}{6pt}}~best to \textcolor[rgb]{0.96,0.78,0.78}{\rule{6pt}{6pt}}~worst within each column.
  \end{minipage}
\end{table}

\subsubsection{Security Validation under Active Attacks}
\label{sec:single-ru-security}

We evaluate \sys under controlled on-path attacks at varying attack rates $p$ (the probability that the adversary perturbs a given synchronization round), shown in Figure~\ref{fig:mitm_replay}.

\noindent\textbf{MITM tampering.}
The adversary modifies packet payloads before they reach the RU. 
Tampered rounds cannot carry valid authentication tags without the undisclosed key; they fail delayed verification and their speculative contributions are deterministically retracted via the correction mechanism. 
RMS offset grows steadily with attack rate as valid updates become sparse and the correction-to-update ratio rises, but one-way delay shows no systematic bias since rejected rounds are excluded from the path-delay estimator by construction.

\noindent\textbf{Replay attacks.}
Replayed bundles carry stale sequence identifiers and are rejected before application by the per-domain monotonicity check, precluding any direct influence on clock state. 
From the servo's perspective, replay is indistinguishable from packet loss: it reduces valid update rate without introducing erroneous inputs. 
RMS offset remains lower than under MITM at comparable attack rates, though variance increases substantially at high rates as update intervals become irregular. One-way delay shows no systematic bias, though variance similarly grows at high attack rates.

Across both attack types, \sys prevents persistent synchronization drift at all tested rates. The contrasting degradation profiles, i.e., correction-induced instability under MITM versus loss-induced variance under replay, reflect the distinct interception points: post-application for integrity failures, pre-application for freshness violations.



\begin{figure}[t]
    \centering
    \includegraphics[width=\linewidth]{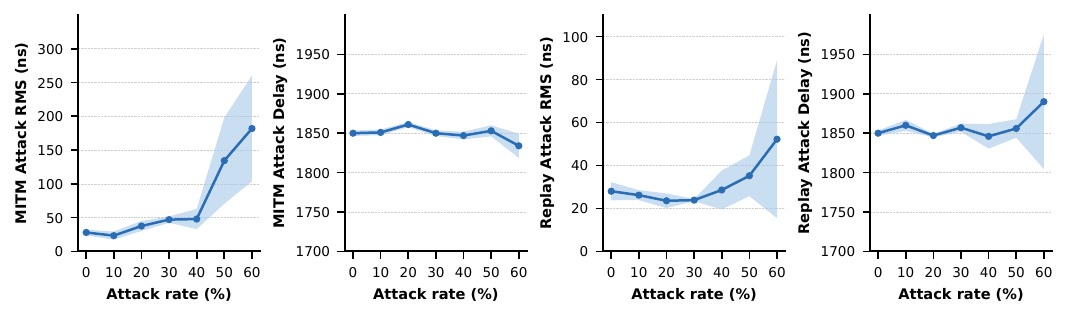}
    \caption{
        Impact of 
        MITM and Replay attacks on \sys under varying attack rates. Shaded regions denote $\pm$ standard deviation.
    }
    \label{fig:mitm_replay}
\end{figure}


\subsection{Multi-RU Scalability}
\label{sec:multi-ru}

We evaluate \sys under a one-to-many topology in which a single DU simultaneously serves two RU slaves within the same L2 multicast domain. As detailed in Section~\ref{sec:setup}, both slaves are emulated via network namespaces on a shared host and rely on software timestamping, which introduces higher baseline jitter than hardware PHC timestamps and accounts for the elevated absolute offset values relative to the single-RU experiments.

Figure~\ref{fig:multiru} reports RMS clock offset and mean one-way delay for both slaves across S1-1 (MACsec-based), S1-3 (ASCON-based), and \sys. Two observations stand out. First, the two slaves exhibit consistent numerical differences across all schemes, which is a result of asymmetric scheduling pressure and unequal resource contention between namespace instances on the shared host, rather than any scheme-specific behavior. Second, within each slave, timing accuracy under \sys is closely comparable to the shared-key baselines: \sys introduces no systematic offset bias or delay inflation when authenticating multiple slaves concurrently. This confirms that the per-round authentication and asynchronous verification path do not add measurable latency to the synchronization control loop even under concurrent multi-slave operation.

Correctness under concurrent operation is preserved through strict per-slave state isolation. Each RU maintains an independent authentication context—comprising its key-chain state, verification buffer, and round indices—and bundles are bound to individual slaves via PTP port identifiers (\texttt{sourcePortIdentity} and \texttt{requestingPortIdentity}), precluding cross-RU state interference. The Delay domain's explicit slave-identity binding (Section~\ref{sec:sys-apply}) further ensures that a \textit{Delay\_Resp} authenticated for one slave cannot be accepted by another, even when both share the same multicast synchronization stream.
Together, these results confirm that \sys scales to concurrent multi-slave deployments without compromising authentication isolation or timing accuracy.

\begin{figure}
    \centering
    \includegraphics[width=0.95\linewidth]{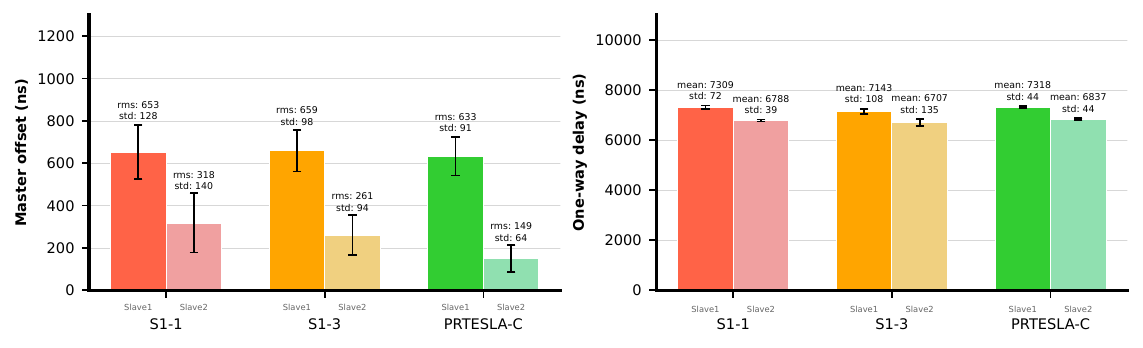}
    \caption{Multi-RU synchronization results with two slaves in the same multicast domain. 
    }
    \label{fig:multiru}
\end{figure}

\subsection{End-to-End System Evaluation}
\label{sec:e2e}

We evaluate \sys in an end-to-end O-RAN deployment based on srsRAN, covering RU--DU initialization, Open Fronthaul (OFH) establishment, UE attachment, and steady-state data transmission. With \sys enabled on the synchronization path, the system initializes successfully: the RU emulator binds to the fronthaul interface, the DU-side gNB completes OFH setup, and the UE attaches and operates normally. No abnormal PTP state transitions, synchronization loss, or RU--DU interface errors are observed, indicating that \sys integrates transparently with the O-RAN control and data planes.

In steady state, \sys preserves the timing behavior of the plain (S0) baseline. In the single-RU configuration, RMS clock offset remains within the same sub-$200$\,ns regime, with comparable frequency-control dynamics and one-way delay statistics. In a dual-RU configuration where two logical RUs synchronize concurrently with a single DU, \sys maintains strict per-RU authentication state and stable operation. RMS offset remains in the same microsecond-level range as the baseline (1.5–1.8\,$\mu$s for S0 and 1.5–2.0\,$\mu$s for \sys), while one-way delay remains comparable (4.4–4.8\,$\mu$s vs. 4.6–5.5\,$\mu$s). The higher absolute values stem from software timestamping and virtualization overhead in the RU emulator environment rather than the authentication mechanism.
Beyond the initial key-chain bootstrap, \sys requires no additional configuration. Overall, these results demonstrate that \sys preserves interoperability and stable synchronization in both single- and multi-RU O-RAN deployments.

\section{Related Works}

\noindent
\textbf{O-RAN architectures and security.}
Polese et al.~\cite{polese2023understanding} and Azariah et al.~\cite{azariah2024survey} provide comprehensive overviews of O-RAN architecture and emphasize the critical role of fronthaul synchronization for 5G performance. Empirical work demonstrates the fragility of this infrastructure: Xing et al.~\cite{xing2024criticality} show that MITM manipulation of fronthaul traffic triggers large-scale service degradation, while Groen et al.~\cite{groen2024securing,groen2024timesafe} demonstrate that PTP spoofing and delay attacks severely disrupt O-RAN timing. Industry analyses identify the open fronthaul as a major attack surface and recommend transport-layer protections such as TLS, IPsec, or MACsec~\cite{quad2023oransecurityreport}. Prior work on O-RAN synchronization-plane security~\cite{maamary2024synchronization, cho2021secure,dik2021transport,dik2023open} largely focuses on generic transport protection without accounting for the impact on tight timing loops. Our work directly targets the synchronization plane, co-designing lightweight authentication with PTP's control loop under O-RAN's strict timing constraints.

\noindent
\textbf{PTP security threats and defenses.}
RFC~7384~\cite{mizrahi2014rfc} identifies PTP threats such as impersonation, replay, and delay attacks; practical studies further show that delay manipulation degrades synchronization in data-center~\cite{decusatis2019impact} and 5G~\cite{alghamdi2021precision} settings. Existing defenses include elliptic-curve signatures~\cite{itkin2017security}, Annex~K-style symmetric authentication~\cite{hirschler2011validation,maftei2018implementing,shereen2019next}, shared-key software authentication~\cite{rezabek2022ptp}, architectural PTP changes~\cite{moussa2016detection}, and PTPsec's path-asymmetry analysis~\cite{finkenzeller2024ptpsec}. However, they typically assume stable pairwise keying, add per-packet overhead, or rely on deployment assumptions ill-suited to O-RAN fronthaul, where multicast distribution, endpoint compromise, and strict timing budgets must be addressed together. Our work provides lightweight, synchronization-aware authentication that preserves tight timing constraints while scaling to broadcast synchronization.


\section{Conclusion}

We address the lack of practical authentication for O-RAN fronthaul synchronization. PRTESLA-C combines per-round delayed disclosure with apply-then-verify-and-correct execution, allowing PTP samples to drive the servo immediately while removing invalid samples from persistent state after verification. Our linuxptp prototype preserves near-baseline synchronization accuracy with low overhead and resists spoofing and replay while bounding the impact of manipulation within the disclosure window. These results show that scalable PTP authentication is feasible under O-RAN fronthaul timing constraints.

\bibliographystyle{plain}
\bibliography{refs}

\begin{thebibliography}{10}

\bibitem{linuxptp}
Linuxptp project.
\newblock \url{https://github.com/richardcochran/linuxptp}, 2026.

\bibitem{ptp4l}
ptp4l linux man page.
\newblock \url{https://linux.die.net/man/8/ptp4l}, 2026.

\bibitem{comp}
3GPP.
\newblock 3gpp tr 36.819 v11.2.0.
\newblock Technical report, 2013.

\bibitem{alghamd2020detection}
Waleed Alghamd and Michael Schukat.
\newblock A detection model against precision time protocol attacks.
\newblock In {\em 2020 3rd International Conference on Computer Applications \& Information Security (ICCAIS)}, pages 1--3. IEEE, 2020.

\bibitem{alghamdi2021precision}
Waleed Alghamdi and Michael Schukat.
\newblock Precision time protocol attack strategies and their resistance to existing security extensions.
\newblock {\em Cybersecurity}, 4(1):12, 2021.

\bibitem{oran_3gpp_split}
O-RAN Alliance.
\newblock O-ran xhaul transport requirements 1.0.
\newblock Technical report, 2021.

\bibitem{oran_planes}
O-RAN Alliance.
\newblock O-ran control, user and synchronization plane specification 17.01.
\newblock Technical report, 2025.

\bibitem{oran_security_req}
O-RAN Alliance.
\newblock O-ran security requirements and controls specifications 11.0.
\newblock Technical report, 2025.

\bibitem{oran_threat}
O-RAN Alliance.
\newblock O-ran security threat modeling and risk assessment.
\newblock Technical report, 2025.

\bibitem{atalay2023securing}
Tolga~O Atalay, Sudip Maitra, Dragoslav Stojadinovic, Angelos Stavrou, and Haining Wang.
\newblock Securing 5g openran with a scalable authorization framework for xapps.
\newblock In {\em IEEE INFOCOM 2023-IEEE Conference on Computer Communications}, pages 1--10. IEEE, 2023.

\bibitem{azariah2024survey}
Wilfrid Azariah, Fransiscus~Asisi Bimo, Chih-Wei Lin, Ray-Guang Cheng, Navid Nikaein, and Rittwik Jana.
\newblock A survey on open radio access networks: Challenges, research directions, and open source approaches.
\newblock {\em Sensors}, 24(3):1038, 2024.

\bibitem{cho2021secure}
Joo~Yeon Cho and Andrew Sergeev.
\newblock Secure open fronthaul interface for 5g networks.
\newblock In {\em Proceedings of the 16th International Conference on Availability, Reliability and Security}, pages 1--6, 2021.

\bibitem{quad2023oransecurityreport}
Quad Critical and Emerging Technology~Working Group.
\newblock Open ran security report.
\newblock 2023.

\bibitem{decusatis2019impact}
Casimer DeCusatis, Robert~M Lynch, William Kluge, John Houston, Paul~A Wojciak, and Steve Guendert.
\newblock Impact of cyberattacks on precision time protocol.
\newblock {\em IEEE Transactions on Instrumentation and Measurement}, 69(5):2172--2181, 2019.

\bibitem{dik2021transport}
Daniel Dik and Michael~St{\"u}bert Berger.
\newblock Transport security considerations for the open-ran fronthaul.
\newblock In {\em 2021 IEEE 4th 5G World Forum (5GWF)}, pages 253--258. IEEE, 2021.

\bibitem{dik2023open}
Daniel Dik and Michael~St{\"u}bert Berger.
\newblock Open-ran fronthaul transport security architecture and implementation.
\newblock {\em IEEE Access}, 11:46185--46203, 2023.

\bibitem{dobraunig2021ascon}
Christoph Dobraunig, Maria Eichlseder, Florian Mendel, and Martin Schl{\"a}ffer.
\newblock Ascon v1. 2: Lightweight authenticated encryption and hashing.
\newblock {\em Journal of Cryptology}, 34(3):33, 2021.

\bibitem{finkenzeller2024ptpsec}
Andreas Finkenzeller, Oliver Butowski, Emanuel Regnath, Mohammad Hamad, and Sebastian Steinhorst.
\newblock Ptpsec: Securing the precision time protocol against time delay attacks using cyclic path asymmetry analysis.
\newblock In {\em IEEE INFOCOM 2024-IEEE Conference on Computer Communications}, pages 461--470. IEEE, 2024.

\bibitem{groen2024timesafe}
Joshua Groen, Simone Di~Valerio, Imtiaz Karim, Davide Villa, Yiewi Zhang, Leonardo Bonati, Michele Polese, Salvatore D'Oro, Tommaso Melodia, Elisa Bertino, et~al.
\newblock Timesafe: Timing interruption monitoring and security assessment for fronthaul environments.
\newblock {\em arXiv preprint arXiv:2412.13049}, 2024.

\bibitem{groen2024securing}
Joshua Groen, Salvatore D'Oro, Utku Demir, Leonardo Bonati, Davide Villa, Michele Polese, Tommaso Melodia, and Kaushik Chowdhury.
\newblock Securing o-ran open interfaces.
\newblock {\em IEEE Transactions on Mobile Computing}, 23(12):11265--11277, 2024.

\bibitem{habibi2021towards}
Mohammad~Asif Habibi, Bin Han, Meysam Nasimi, Nandish~P Kuruvatti, Amina Fellan, and Hans~D Schotten.
\newblock Towards a fully virtualized, cloudified, and slicing-aware ran for 6g mobile networks.
\newblock In {\em 6G Mobile Wireless Networks}, pages 327--358. Springer, 2021.

\bibitem{hirschler2011validation}
Bernd Hirschler and Albert Treytl.
\newblock Validation and verification of ieee 1588 annex k.
\newblock In {\em 2011 IEEE International Symposium on Precision Clock Synchronization for Measurement, Control and Communication}, pages 44--49. IEEE, 2011.

\bibitem{hung2024security}
Cheng-Feng Hung, You-Run Chen, Chi-Heng Tseng, and Shin-Ming Cheng.
\newblock Security threats to xapps access control and e2 interface in o-ran.
\newblock {\em IEEE Open Journal of the Communications Society}, 5:1197--1203, 2024.

\bibitem{itkin2017security}
Eyal Itkin and Avishai Wool.
\newblock A security analysis and revised security extension for the precision time protocol.
\newblock {\em IEEE Transactions on Dependable and Secure Computing}, 17(1):22--34, 2017.

\bibitem{janzen2024oh}
Leon Janzen, Lucas Becker, Colin Wiesen{\"a}cker, and Matthias Hollick.
\newblock Oh no, my $\{$RAN$\}$! breaking into an $\{$O-RAN$\}$ 5g indoor base station.
\newblock In {\em 18th USENIX WOOT Conference on Offensive Technologies (WOOT 24)}, pages 101--115, 2024.

\bibitem{klement2024securing}
Felix Klement, Alessandro Brighente, Michele Polese, Mauro Conti, and Stefan Katzenbeisser.
\newblock Securing the open ran infrastructure: Exploring vulnerabilities in kubernetes deployments.
\newblock In {\em 2024 IEEE 10th International Conference on Network Softwarization (NetSoft)}, pages 185--189. IEEE, 2024.

\bibitem{maamary2024synchronization}
Assrar Maamary, Hyame~Assem Alameddine, Mourad Debbabi, and Chadi Assi.
\newblock Synchronization plane in o-ran: Overview, security and research directions.
\newblock {\em IEEE Communications Magazine}, 63(2):88--94, 2024.

\bibitem{maftei2018implementing}
Dragos Maftei, Radim Bartos, Bob Noseworthy, and Timothy Carlin.
\newblock Implementing proposed ieee 1588 integrated security mechanism.
\newblock In {\em 2018 IEEE International Symposium on Precision Clock Synchronization for Measurement, Control, and Communication (ISPCS)}, pages 1--6. IEEE, 2018.

\bibitem{mizrahi2014rfc}
T~Mizrahi.
\newblock Rfc 7384: Security requirements of time protocols in packet switched networks, 2014.

\bibitem{mizrahi2011time}
Tal Mizrahi.
\newblock Time synchronization security using ipsec and macsec.
\newblock In {\em 2011 IEEE International Symposium on Precision Clock Synchronization for Measurement, Control and Communication}, pages 38--43. IEEE, 2011.

\bibitem{moussa2015detection}
Bassam Moussa, Mourad Debbabi, and Chadi Assi.
\newblock A detection and mitigation model for ptp delay attack in a smart grid substation.
\newblock In {\em 2015 IEEE International Conference on Smart Grid Communications (SmartGridComm)}, pages 497--502. IEEE, 2015.

\bibitem{moussa2016detection}
Bassam Moussa, Mourad Debbabi, and Chadi Assi.
\newblock A detection and mitigation model for ptp delay attack in an iec 61850 substation.
\newblock {\em IEEE Transactions on Smart Grid}, 9(5):3954--3965, 2016.

\bibitem{moussa2018securing}
Bassam Moussa, Chantale Robillard, Alf Zugenmaier, Marthe Kassouf, Mourad Debbabi, and Chadi Assi.
\newblock Securing the precision time protocol (ptp) against fake timestamps.
\newblock {\em IEEE communications letters}, 23(2):278--281, 2018.

\bibitem{municio2023ran}
Esteban Municio, Gines Garcia-Aviles, Andres Garcia-Saavedra, and Xavier Costa-P{\'e}rez.
\newblock O-ran: Analysis of latency-critical interfaces and overview of time sensitive networking solutions.
\newblock {\em IEEE Communications Standards Magazine}, 7(3):82--89, 2023.

\bibitem{nokia_oran_sec}
Nokia.
\newblock Open ran security.
\newblock Technical report, 2022.

\bibitem{perrig2003tesla}
Adrian Perrig, J~Doug Tygar, Adrian Perrig, and JD~Tygar.
\newblock Tesla broadcast authentication.
\newblock {\em Secure Broadcast Communication: In Wired and Wireless Networks}, pages 29--53, 2003.

\bibitem{polese2023understanding}
Michele Polese, Leonardo Bonati, Salvatore D’oro, Stefano Basagni, and Tommaso Melodia.
\newblock Understanding o-ran: Architecture, interfaces, algorithms, security, and research challenges.
\newblock {\em IEEE Communications Surveys \& Tutorials}, 25(2):1376--1411, 2023.

\bibitem{rezabek2022ptp}
Filip Rezabek, Max Helm, Tizian Leonhardt, and Georg Carle.
\newblock Ptp security measures and their impact on synchronization accuracy.
\newblock In {\em 2022 18th International Conference on Network and Service Management (CNSM)}, pages 109--117. IEEE, 2022.

\bibitem{shereen2019next}
Ezzeldin Shereen, Florian Bitard, Gy{\"o}rgy D{\'a}n, Tolga Sel, and Steffen Fries.
\newblock Next steps in security for time synchronization: Experiences from implementing ieee 1588 v2. 1.
\newblock In {\em 2019 IEEE International Symposium on Precision Clock Synchronization for Measurement, Control, and Communication (ISPCS)}, pages 1--6. IEEE, 2019.

\bibitem{ptp1}
IEC/IEEE~International Standard.
\newblock Precision clock synchronization protocol for networked measurement and control systems.
\newblock Technical report, 2021.

\bibitem{thimmaraju2024security}
Kashyap Thimmaraju, Altaf Shaik, Sunniva Fl{\"u}ck, Pere Joan~Fullana Mora, Christian Werling, and Jean-Pierre Seifert.
\newblock Security testing the o-ran near-real time ric \& a1 interface.
\newblock In {\em Proceedings of the 17th ACM Conference on Security and Privacy in Wireless and Mobile Networks}, pages 277--287, 2024.

\bibitem{xing2024criticality}
Jiarong Xing, Sophia Yoo, Xenofon Foukas, Daehyeok Kim, and Michael~K Reiter.
\newblock On the criticality of integrity protection in 5g fronthaul networks.
\newblock In {\em 33rd USENIX Security Symposium (USENIX Security 24)}, pages 4463--4479, 2024.

\end{thebibliography}

\appendix

\section{Appendices}
\subsection{Protocol Details}


\begin{algorithm*}[t]
\caption{\sys Protocol Flow (Per-Round TESLA with Correction)}
\label{alg:sys-flow}
\small
\raggedright
\textbf{Parameters:} disclosure delay $d$ (rounds); round interval $T_{\text{int}}$; per-epoch chain length $N$; preannounce margin $R$ (rounds); verification deadline $W$ (Section~4.3); optional slew bound $S_{\max}$.\\
\textbf{Crypto:} one-way function $F$; MAC-key derivation $F'$; per-round MAC $\textsf{ASCON}(\cdot)$.\\
\textbf{DU state:} epoch id $e$, round counter $i$, current chain $\{K_i\}$ with public anchor $K_0$; next-epoch chain $\{K'_j\}$ and anchor $K'_0$ when preannouncing.\\
\textbf{RU state:} expected round id $\textit{SeqExp}$; per-round buffer $\mathcal{B}[\cdot]$ for (\textit{Sync}, \textit{Follow\_Up}, \textit{Delay\_Req}, \textit{Delay\_Resp}, $\textit{MAC}$, $\textit{RolloverMeta}$); ledger $\mathcal{L}[\cdot]$; last accepted disclosed key $\texttt{Klast}$ and its index $\texttt{Ilast}$; cached next anchor $\texttt{NextCommit}$ and switch round $\texttt{SwitchPoint}$ (optional).\\
\hrulefill

\begin{algorithmic}[1]
\item \textbf{Bootstrap (minimally trusted).} DU provisions each RU with $(e,\, i_0,\, d,\, T_{\text{int}},\, K_0)$ and initializes $i\!\leftarrow\! i_0$. RU sets $\textit{SeqExp}\!\leftarrow\! i_0$, $\texttt{Klast}\!\leftarrow\! K_0$, $\texttt{Ilast}\!\leftarrow\!0$.

\item \textbf{Per-round message exchange.} For each round $i$, DU and RU execute standard PTP (\textit{Sync}, \textit{Follow\_Up}, \textit{Delay\_Req}, \textit{Delay\_Resp}). RU buffers received \textit{Sync}/\textit{Follow\_Up}/\textit{Delay\_Req} into $\mathcal{B}[i]$ using PTP identities to associate messages with the correct slave instance.

\item \textbf{DU: compute tag and disclose key.}
Upon generating \textit{Delay\_Resp} for round $i$, DU forms the authenticated bundle
$M_i^{\text{bundle}}$ and sets $\textit{SeqNum}_i \!\leftarrow\! i$.
Let $\textit{RolloverMeta}_i$ be empty unless preannouncement is active (Step~4).
DU computes the per-round tag:
\[
\text{MAC}_i \gets \textsf{ASCON}\!\left(F'(K_i),\; M_i^{\text{bundle}} \,\Vert\, \textit{SeqNum}_i \,\Vert\, \textit{RolloverMeta}_i\right).
\]
DU attaches $(\textit{SeqNum}_i,\, \text{MAC}_i)$ to \textit{Delay\_Resp}, and discloses $K_{i-d}$ (if $i\ge d$) in the same message.

\item \textbf{DU: epoch preannouncement (optional).}
If the remaining rounds in the current epoch are $\le R$ and the next epoch is not yet announced, DU generates a fresh next chain $\{K'_{N},\ldots,K'_0\}$ and sets $\texttt{NextCommit}\!\leftarrow\!K'_0$ and $\texttt{SwitchPoint}\!\leftarrow\! i_{\text{sw}}$ (a future round index). For rounds during preannouncement, DU sets
$\textit{RolloverMeta}_i \gets (\texttt{NextCommit}, \texttt{SwitchPoint})$; otherwise $\textit{RolloverMeta}_i \gets \emptyset$.

\item \textbf{RU: admission checks and buffering (pre-apply).}
Upon receiving \textit{Delay\_Resp} carrying $(\textit{SeqNum}_i,\text{MAC}_i,K_{i-d},\textit{RolloverMeta}_i)$, RU first enforces freshness by checking $\textit{SeqNum}_i \ge \textit{SeqExp}$ and discarding stale responses (replay) before application. If accepted, RU stores the response and TLV fields into $\mathcal{B}[i]$ and updates $\textit{SeqExp}\!\leftarrow\! i+1$. If $\textit{RolloverMeta}_i\neq\emptyset$, RU caches $(\texttt{NextCommit},\texttt{SwitchPoint})$.

\item \textbf{RU: apply fast path (speculative).}
If $\mathcal{B}[i]$ contains a complete four-message round, RU immediately computes offset/delay and applies the servo update (optionally bounded by $S_{\max}$). RU records the minimal incremental state in ledger $\mathcal{L}[i]$ and sets a verification deadline $\texttt{deadline}[i]\!\leftarrow\!\texttt{now}+W$.

\item \textbf{RU: key-chain validation and delayed verification.}
If $i\ge d$, RU validates the disclosed key $K_{i-d}$ against the current epoch anchor. Concretely, RU checks that $K_{i-d}$ is consistent with the one-way chain and the last accepted disclosure (gap-tolerant), e.g.,
$F^{\Delta}(K_{i-d}) = \texttt{Klast}$ for some expected $\Delta$ derived from indices.
On success, RU updates $(\texttt{Klast},\texttt{Ilast})$ and verifies round $(i-d)$ by recomputing
\[
\quad \quad \text{MAC}'_{i-d} \gets \textsf{ASCON}\!\left(F'(K_{i-d}),\; M_{i-d}^{\text{bundle}} \,\Vert\, \textit{SeqNum}_{i-d} \,\Vert\, \textit{RolloverMeta}_{i-d}\right),
\]
and comparing it with the stored $\text{MAC}_{i-d}$.

\item \textbf{RU: commit on success; correct on failure/timeout.}
Upon verifying round $(i-d)$, RU resolves its speculative effects using the per-round ledger $\mathcal{L}[i-d]$:
if verification succeeds, RU \emph{commits} the round by marking it authenticated, discarding the buffered bundle $\mathcal{B}[i-d]$, and releasing the ledger entry $\mathcal{L}[i-d]$ (the already-applied update remains in effect).
Otherwise (MAC mismatch, invalid disclosure, missing bundle components, or verification deadline $W$ expired), RU \emph{rejects} the round and performs a deterministic control-level correction without rolling back wall-clock time. Concretely, RU (i) cancels the offset/servo input applied for round $(i-d)$ and reverts the corresponding controller-state increment recorded in $\mathcal{L}[i-d]$, and (ii) subtracts the recorded incremental update to the path-delay estimator to restore it to its pre-round value (without resetting filter history). RU then marks the round as rejected/unverifiable and frees $\mathcal{B}[i-d]$ and $\mathcal{L}[i-d]$. 

\item \textbf{RU: epoch switch at \texttt{SwitchPoint}.}
If valid rollover metadata $(\texttt{NextCommit}, \texttt{SwitchPoint})$ has been cached, RU activates the new epoch when the local round index reaches \texttt{SwitchPoint}.
Specifically, RU sets the current verification anchor for newly arriving rounds to $\texttt{NextCommit}$ (the next-chain commitment $K'_0$) and advances the epoch identifier, while continuing to verify any previously buffered rounds using the old epoch key chain.
To support the delayed-authentication pipeline, RU retains the previous epoch
verification state until all pending rounds whose authentication keys belong to the old chain (at most $d$ rounds) have been verified or resolved.
After these rounds are completed, the old epoch state can be safely discarded.
If \texttt{NextCommit} is missing or fails validation, RU enters holdover and rejects new-epoch synchronization updates until a valid commitment is re-provisioned.


\end{algorithmic}
\end{algorithm*}

\end{document}